\pdfoutput=1
\documentclass[sigconf]{acmart}
\usepackage{algorithm}
\usepackage{algorithmicx}
\usepackage{algpseudocode}
\usepackage{graphicx}
\usepackage{caption}
\usepackage{subcaption}

\AtBeginDocument{%
  \providecommand\BibTeX{{%
    \normalfont B\kern-0.5em{\scshape i\kern-0.25em b}\kern-0.8em\TeX}}}

\copyrightyear{2020} 
\acmYear{2020} 
\setcopyright{acmlicensed}\acmConference[e-Energy'20]{The Eleventh ACM International Conference on Future Energy Systems}{June 22--26, 2020}{Virtual Event, Australia}
\acmBooktitle{The Eleventh ACM International Conference on Future Energy Systems (e-Energy'20), June 22--26, 2020, Virtual Event, Australia}
\acmPrice{15.00}
\acmDOI{10.1145/3396851.3402369}
\acmISBN{978-1-4503-8009-6/20/06}

\begin{document}
\emergencystretch 3em

\title[e-Energy’20 AMLIES]{Optimizing carbon tax for decentralized electricity markets using an agent-based model}

\author{Alexander J. M. Kell, A. Stephen McGough, Matthew Forshaw}
\affiliation{%
  \department{School of Computing}
  \institution{Newcastle University}
  \city{Newcastle upon Tyne}
  \country{UK}
}
\email{{a.kell2, stephen.mcgough, matthew.forshaw}@newcastle.ac.uk}

%
\renewcommand{\shortauthors}{Kell et al.}


\begin{abstract}
 

Averting the effects of anthropogenic climate change requires a transition from fossil fuels to low-carbon technology. A way to achieve this is to decarbonize the electricity grid. However, further efforts must be made in other fields such as transport and heating for full decarbonization. This would reduce carbon emissions due to electricity generation, and also help to decarbonize other sources such as automotive and heating by enabling a low-carbon alternative. Carbon taxes have been shown to be an efficient way to aid in this transition.

In this paper, we demonstrate how to to find optimal carbon tax policies through a genetic algorithm approach, using the electricity market agent-based model ElecSim. To achieve this, we use the NSGA-II genetic algorithm to minimize average electricity price and relative carbon intensity of the electricity mix. We demonstrate that it is possible to find a range of carbon taxes to suit differing objectives. 

Our results show that we are able to minimize electricity cost to below \textsterling10/MWh as well as carbon intensity to zero in every case. In terms of the optimal carbon tax strategy, we found that an increasing strategy between 2020 and 2035 was preferable. Each of the Pareto-front optimal tax strategies are at least above \textsterling81/tCO2 for every year. The mean carbon tax strategy was \textsterling240/tCO2.

\end{abstract}




\maketitle

\section{Introduction}


Computer simulation allows practitioners to model real-world systems using software. These simulations allow for `\textit{what-if} ' analyses which can provide an indication as to how a system may behave under specific policies, environments and assumptions. These simulations become particularly important in real systems which have high costs, impacts or risks associated with them. A `digital twin' is a concept which has emerged in recent years. This is defined as a simulation of a specific instance of a system. This digital twin can then be used to learn of optimization techniques that can be applied to the real system. This foregoes the necessity of experimenting with the real system, avoiding potential adverse side effects.


An electricity market is an example of a complex system which can be modelled using a digital twin. Disruptions to electricity supply, a substantial increase in the cost of electricity or unrestrained carbon emissions have the potential to destabilize economies~\cite{Kaseke2013,Masson-Delmotte2018}. It is for reasons such as these that electricity market models are used to test hypotheses, develop strategies and gain an understanding of underlying dynamics to prevent undesirable consequences \cite{Jebaraj2006}. 

In this paper, we use the electricity market agent-based model ElecSim to find an optimum carbon tax policy \cite{Kell}. Specifically, we use a genetic algorithm to find a carbon tax policy to reduce both average electricity price and the relative carbon density by 2035 for the UK electricity market. 

Carbon taxes have been shown to quickly lower emissions and lower the costs to the public \cite{Wittneben2009}. Carbon taxes are able to send clear price signals, as opposed to a cap-and-trade scheme, such as the EU Emissions Trading System, which has shown to be unstable~\cite{Wittneben2009}.

In this paper, we use the reference scenario projected by the UK Government's Department for Business \& Industrial Strategy (BEIS) with model parameters calibrated by Kell \textit{et al.} \cite{DBEIS2019,Kell2020}. This reference scenario projects energy and emissions until 2035. We undertake various carbon tax policy interventions to see how we could reduce relative carbon density whilst at the same time, reduce the average electricity price.

The parameter space we optimize over is the carbon tax price over a 17 year period from 2018 to 2035. The carbon price is used to influence the objectives of average electricity price and relative carbon intensity in 2035. Grid and random search are approaches which trial parameters at evenly distributed spaces and random spaces respectively. These approaches are often inefficient, however, and require an increased number of simulations due to their static nature. Genetic Algorithms, in contrast, use an evolutionary computing approach to find global optimal solutions faster.


In order to optimize over two potentially competing objectives, i.e. average electricity price and relative carbon intensity, we use the Non-Dominated Sorting Genetic Algorithm II (NSGA-II) \cite{Valkanas2014}. The NSGA-II algorithm can approximate a Pareto frontier ~\cite{Pareto1927, Stadler1979}. A Pareto frontier is a curve in which there is no solution which is better than another along the curve for different sets of parameters. In this context, better means that a solution is closer to the optimal for a particular combination of objectives.

We find that the rewards of average electricity price and relative carbon intensity are not mutually exclusive. That is, it is possible to have both a lower average electricity price and a lower relative carbon price. This is due to the low short-run marginal cost of renewable technology, which has been shown to lower electricity prices \cite{OMahoney2011}.

The main contribution of this paper is to explore carbon tax strategies using genetic algorithms for multi-objective optimization. 

The following sections are set out as follows. Section \ref{sec:lit_review} covers examples of optimisations using genetic algorithms and different carbon strategies. Section \ref{sec:optimization_methods} details the optimization techniques applied. Section \ref{sec:sim_environment} explores the electricity market model used. We present our results in Section \ref{sec:results}, and conclude in Section \ref{sec:conclusion}.








\section{Literature Review}
\label{sec:lit_review}

Multi-objective optimization problems are commonplace. In this section, we review multiple applications that have used multi-objective optimization, as well as explore the literature which focus on finding optimal carbon tax strategies.

\subsection{Examples of Optimization}

Similar to our work, Ascione  \textit{et al}. use the NSGA-II algorithm to generate a Pareto front to optimize for two objectives: operating cost for space conditioning and thermal comfort \cite{Ascione2016}. The aim of their paper is to optimize the hourly set point temperatures with a day-ahead planning horizon. A Pareto front is generated, which allows a user to select a solution according to their comfort needs and economic constraints. This work showed a reduction in operating costs by up to 56\% as well as improved thermal comfort.

Gorza\l{}czany \textit{et al}. also apply the NSGA-II algorithm. However, they apply it to the credit classification problem \cite{Gorzaczany2016}. The objectives optimized over were accuracy and interpretability when making financial decisions such as credit scoring and bankruptcy prediction. This technique was able to significantly outperform the alternative methods in terms of interpretability while remaining competitive or superior in terms of the accuracy and speed of decision making in comparison with the existing classification methods.


\subsection{Carbon Tax Strategies}

In this section, we explore different strategies employed in the literature to analyze the benefits and consequences of a carbon tax. To the best of our knowledge, we are the first to employ a multi-objective optimization algorithm to minimize average electricity price and relative carbon density.

Levin \textit{et al}. use an optimization model to analyze market and investment impacts of several incentive mechanisms to support investment in renewable energy and carbon emission reductions~\cite{Levin2019}. Carbon tax was found to be the most cost-efficient method of reducing emissions.

Zhou \textit{et al.} construct a social welfare model based on a Stackelberg game~\cite{Zhou2019}. The differences and similarities between a flat carbon tax and an increasing block tariff carbon tax are analyzed using a numerical simulation. This work shows that an increasing block tariff carbon tax policy can significantly reduce tax burdens for manufacturers and encourage low-carbon production. In contrast to Zhou \textit{et al}. we trial multiple different carbon tax strategies using a machine learning approach. 

Li \textit{et al}. use a hierarchical carbon market scheduling model to reduce carbon emissions \cite{Li2017}. Multi-objective optimization was applied to discover optimal behaviours for policymakers, customers and generators to minimize the costs incurred by these actors. Our work, however, focuses on the different strategies of carbon tax as opposed to optimal actor behaviour.

\section{Optimization methods}
\label{sec:optimization_methods}

Multi-objective optimization allows practitioners to overcome the problems with optimizing multiple objectives with classical optimization techniques. Multi-objective optimization algorithms are able to generate Pareto-optimal solutions as opposed to converting the multiple objectives into a single-objective problem. A single-objective problem assumes that there is only a single optimum, and that other combinations are inferior. This may not be the case, as different solutions are superior for a different set of circumstances. A Pareto frontier is made up of many Pareto-optimal solutions which can be displayed graphically. A user is then able to choose between various solutions and trade-offs according to their wishes.

The NSGA-II algorithm, a multi-objective genetic optimization algorithm, is able to generate a Pareto frontier in a single optimization run. 

In the following sub-sections, we detail the standard genetic algorithm followed by the NSGA-II algorithm.

\subsection{Genetic Algorithms}

Genetic Algorithms (GAs) ~\cite{Holland1975} are a class of evolutionary algorithm which can be used for optimization. 

As shown by Algorithm \ref{genetic-algorithm}, initially, a population of parameters $P_{0}$ are generated to be trialled in the simulation. $P_{0}$ is then evaluated for fitness, by running an instance of the simulation with the respective parameters as inputs. Where fitness in this case is the reward that is to be optimised. Next, a subset of individuals from $P_{0}$ are chosen for mating, $C_{t+1} \subset P_{t}$. This subset of individuals are selected proportional to their fitness. For mating with the subset $C_{t+1}$, the `fitter' individuals have a higher chance of reproducing to create the offspring group $C'_{t+1}$. The individuals of $C'_{t+1}$ have characteristics dependent on the genetic operators, crossover and mutation \cite{mitchell1998introduction}. These genetic operators are an implementation decision \cite{FogelDavidB2009}. The new population $P_{t+1}$ is then created by merging individuals from $C'_{t+1}$ and $P_{t}$. See Algorithm \ref{genetic-algorithm} for detailed pseudocode.
\begin{algorithm}[t]
\begin{algorithmic}[1]
\State $t=0$
\State initialize $P_{t}$
\State evaluate structures in $P_{t}$
\While {termination condition not satisfied}
\State $t=t+1$
\State select reproduction $C_{t}$ from $P_{t-1}$
\State recombine and mutate structures in $C_{t}$

forming $C'_{t}$
\State evaluate structures in $C'_{t}$
\State select each individual for $P_{t}$ from $C'_{t}$ 

or $P_{t-1}$
\EndWhile
\caption{Genetic algorithm \cite{FogelDavidB2009}}
\label{genetic-algorithm}
\end{algorithmic}
\end{algorithm}


\subsection{NSGA-II}

NSGA-II is efficient for multi-objective optimization on a number of benchmark problems and finds a better spread of solutions than Pareto Archived Evolution Strategy (PAES)~\cite{Knowles1999} and Strength Pareto EA (SPEA)~\cite{Zitzler2006} when approximating the true Pareto-optimal front \cite{Valkanas2014}.

The majority of multi-objective optimization algorithms use the concept of \emph{domination} during population selection \cite{Burke2014}. A non-dominated algorithm, however, seeks to achieve the Pareto-optimal solution. This is where no single solution should dominate another. An individual solution $\mathbf{x}^{1}$ is said to dominate another $\mathbf{x}^{2}$, if and only if there is no objective of $\mathbf{x}^{1}$ that is worse than objective of $\mathbf{x}^{2}$ and at least one objective of $\mathbf{x}^{1}$ is better than the same objective of $\mathbf{x}^{2}$ \cite{Bao2017}.

Non-domination sorting is the process of finding a set of solutions which do not dominate each other and make up the Pareto front. See Figure \ref{fig:pareto-layering}a for a visual representation, where $f_1$ and $f_2$ are two objectives to minimize and L1, L2 and L3 are dominated layers.

In this section, we define the processes used by the NSGA-II algorithm to determine which solutions to keep:
\subsubsection{Non-dominated sorting}
We assume that there are $M$ objective functions to minimise, and that ${\bf x^{1}} = \{x_j^{1}\}$ and $\bf x^{2}$ are two solutions. $x_j^{1}<x_j^{2}$ implies solution $\bf x^{1}$ is better than solution $\bf x^{2}$ on objective $j$. A solution $\mathbf{x}^{1}$ is said to dominate the solution $\mathbf{x}^{2}$ if the following conditions are true:
\begin{enumerate}
  \item The solution $\mathbf{x}^{1}$ is no worse than $\mathbf{x}^{2}$ in every objective. I.e. $x^{1}_j \leq x^{2}_j \;\;  \forall j \in\{1,2,\ldots,M\}$.
  \item The solution $\mathbf{x}^{1}$ is better than $\mathbf{x}^{2}$ in at least one objective. I.e. $\exists\  {j}\in \{ 1,2,\ldots,M\} \;\; s.t. \;\;x^{1}_j < x^{2}_j$.
\end{enumerate}

\begin{figure}[t] 
  \vskip -10pt
  \center
  \includegraphics[width=0.2\textwidth]{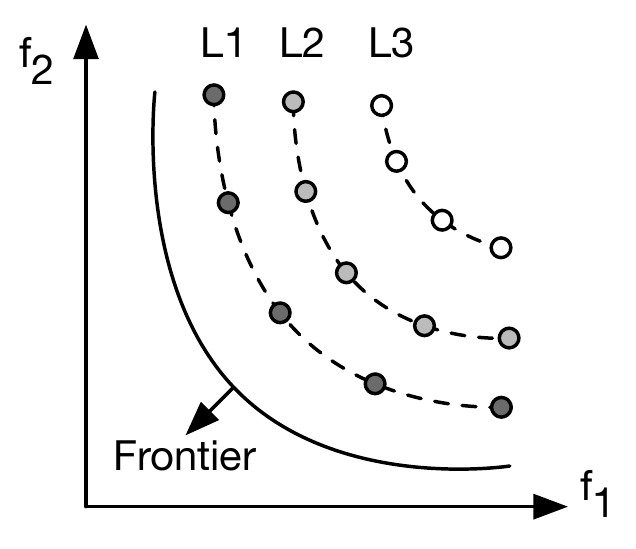}
  \includegraphics[width=0.270\textwidth]{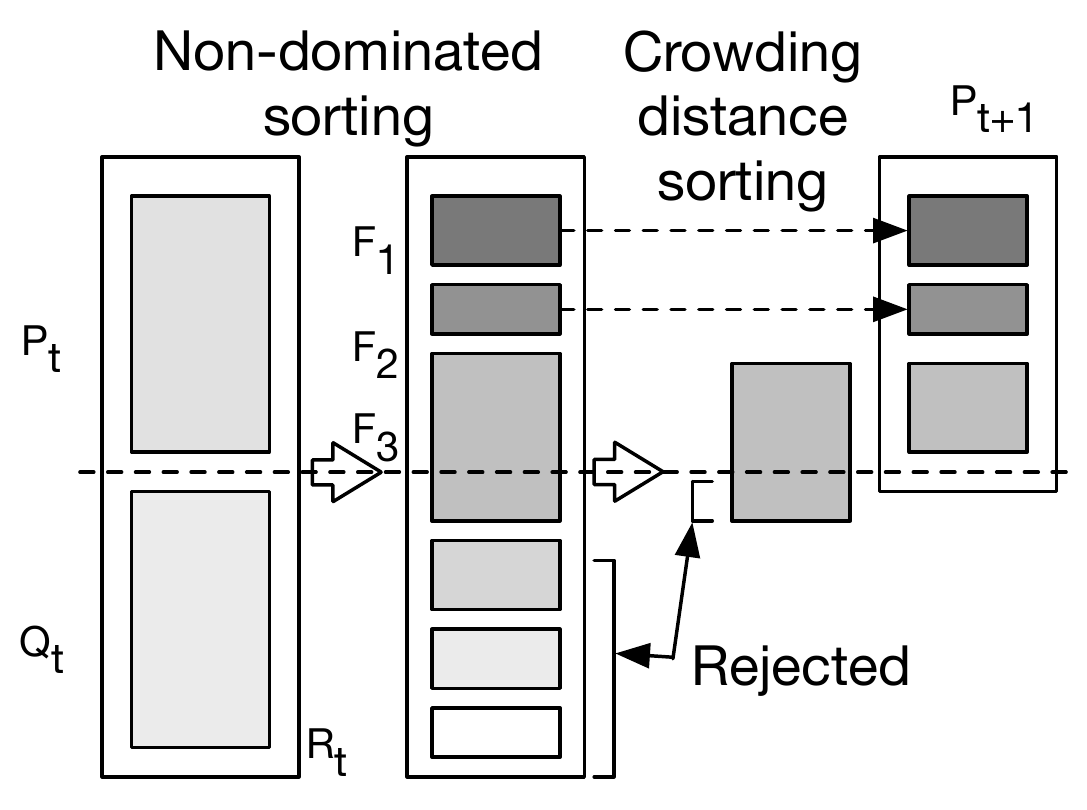}
  \vskip -8pt
  \caption{a) Schematic of non-dominated sorting with solution layering b) Schematic of the NSGA-II procedure}
  \label{fig:pareto-layering}
  \vskip -15pt
\end{figure}

Next, each of the solutions are ranked according to their level of non-domination. An example of this ranking is shown in Figure \ref{fig:pareto-layering}a. Here, $f_1$ and $f_2$ are the objectives to be minimised. The Pareto front is the first front. All of the solutions in the Pareto front are not dominated by any other solution. The solutions in layer 1, L1, are dominated only by those in the Pareto front, and are non-dominated by those in L2 and L3.

The solutions are then ranked according to their layer. For example, the solutions in the Pareto front are given a fitness rank ($i_{rank}$) of 1, solutions in L1 have an $i_{rank}$ of 2.

\subsubsection{Density Estimation}
($i_{distance}$) is calculated for each solution. This is the average distance between the two closest points to the solution in question.

\subsubsection{Crowded comparison operator}
($\prec_n$) is used to ensure that the final frontier is an evenly spread out Pareto-optimal front. This is achieved by using the two attributes: $(i_{rank})$ and$(i_{distance})$. 
The partial order is then defined as:\\    
$i\prec_nj$ if $(i_{rank}<j_{rank})$ or $((i_{rank}=j_{rank})$ and  $(i_{distance}>j_{distance}))$ \cite{Valkanas2014}.

This order prefers solutions with a lower rank $i_{rank}$. For solutions with the same rank, the solution in the less dense area is preferred.

\subsubsection{Main loop}

Similarly to the standard GA, a population $P_{0}$ is created with random parameters. The solutions of $P_0$ are then sorted according to non-domination. The child population $C'_{1}$ of size $N$ is then created by binary tournament selection, recombination and mutation operators. In this case, tournament selection is the process of evaluating and comparing the fitness of various individuals within a population. In binary tournament selection, two individuals are chosen at random, the fitnesses are evaluated, and the individual with the better solution is selected~\cite{AbdRahman2016}.

\begin{algorithm}[b]
\begin{algorithmic}[1]
\State $R_t=P_t \cup C'_t$ combine parent and child population
\State $\mathcal{F} = $ fast-non-dominated-sort $(R_t)$ 

where $\mathcal{F}=(\mathcal{F}_1, \mathcal{F}_2,\ldots)$
\State $P_{t+1}=\emptyset$
\While $\left|P_{t+1}<N\right|$
\State Calculate the crowding distance of $(\mathcal{F}_i)$)
\State $P_{t+1}=P_{t+1}\cup \mathcal{F}_i$
\EndWhile
\State Sort($P_{t+1}, \prec_n$) sort in descending order using $\prec_n$
\State $P_{t+1} = P_{t+1}[0:N]$ select the first $N$ elements of $P_{t+1}$
\State $Q_{t+1} = $ make-new-population$(P_{t+1})$ using 

selection, crossover and mutation to create 

the new population $Q_{t+1}$
\State $t=t+1$
\caption{NSGA-II main loop \cite{Valkanas2014}}
\label{algo:nsga2}
\end{algorithmic}
\end{algorithm}

Next, a new combined population is formed $R_{t}=P_{t} \cup C'_{t}$. $R_t$ has a size of $2N$. $R_t$ is then sorted according to non-domination. A new population is then formed $P_{t+1}$. Solutions are added from the sorted $R_t$ in order of non-domination. Solutions are added until the size of $P_{t+1}$ exceeds $N$. The solutions from the last layer are prioritised based on having the largest crowding distance~\cite{Valkanas2014}.

This process is shown in Figure \ref{fig:pareto-layering}b, which is repeated until the termination condition is met. A termination condition could be:  no significant improvement over $X$ iterations or a specified number of iterations have been performed. The full procedure is detailed formally by Algorithm \ref{algo:nsga2}.

\subsection{Carbon Optimization Application}

In this section, we describe how the genetic algorithm is applied in our carbon optimization case. We use multi-objective optimization to find a solution which has both a low carbon emission and low average electricity price. The parameters that we adjust is the carbon tax between the years 2018 and 2035.

The mating steps work by, initially, taking the sets of carbon prices over the 17 year period (2018 to 2035) which have the best rewards (lowest relative carbon emissions and average electricity price). These carbon prices are then mated with a probability of 90\%, creating child carbon prices. The children are mutated with a probability of 5\%. Therefore, 5\% of children have a carbon price which is not inherited from the parents. Over time, the mutations and inherited properties tend to a population with more desirable rewards.

\section{Experimental Setup}
\label{sec:sim_environment}

\subsection{Simulation Environment}
In order to evaluate the different carbon strategies, we used the model developed by Kell \textit{et al}., ElecSim \cite{Kell,Kell2020}. ElecSim is an agent-based model which mimics the behaviour of decentralized electricity markets. For this paper, we parametrized the model to data for the UK in 2018 to act as a digital twin of the UK electricity market. This includes the power plants in operation in 2018, and the funds available to their respective companies \cite{dukes_511, companies_house}. ElecSim is validated by being instantiated by data from 2013 and projected forward to 2018, with a mean absolute scaled error (MASE) below or equal to 0.701 for all generator types \cite{Kell2020}. 

\begin{figure}
\centering
\includegraphics[width=0.44\textwidth]{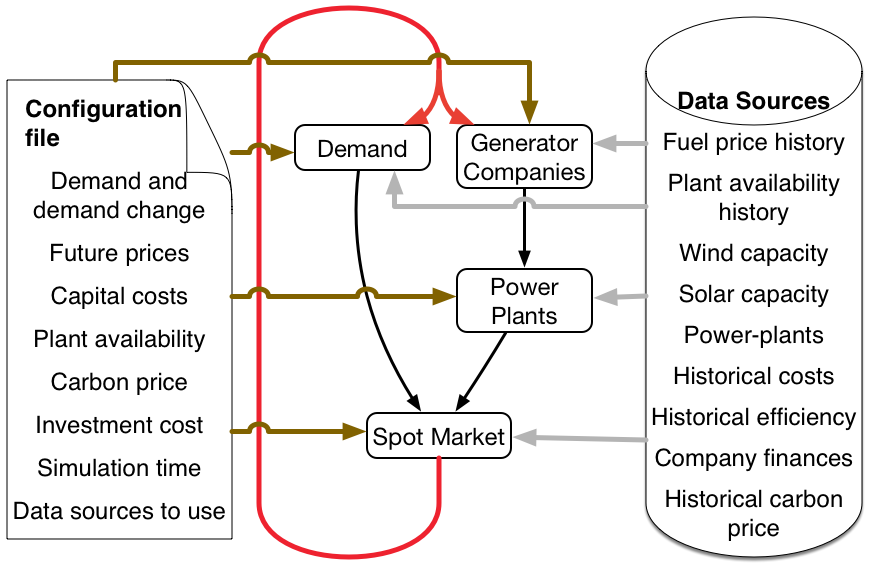}
\caption{System overview of ElecSim \cite{Kell}.}
\label{fig:model_details}
\end{figure}

Six fundamental sections make up ElecSim: 1) power plant data; 2) scenario data; 3) the time-steps of the algorithm; 4) the power exchange; 5) the investment algorithm and 6) the generation companies (GenCos) as agents. ElecSim uses a subset of representative days of electricity demand, solar irradiance and wind speed to approximate a full year. In this context, representative days are a subset of days which, when scaled up, can adequately represent a year. Figure \ref{fig:model_details} details how these components interact. 

Specifically, the configuration file details the scenario which can be set by the user. This includes electricity demand, carbon price and fuel prices. The data sources parametrize the digital twin to a particular country, including information such as wind capacity and power plants in operation. Generation Companies own and invest in power plants. These power plants are then matched to electricity demand using a spot market.

The market runs in merit-order dispatch and bids are made by the power plant's short-run marginal cost (SRMC). Investment in power plants are based upon a net present value (NPV) calculation. Decisions are made every year, where new generators enter the market. NPV is able to evaluate and compare investments with cash flows spread over many years. This is shown formally in Equation \ref{eq:npv_eq}, where $t$ is the year of the cash flow, $i$ is the discount rate, $N$ is the total number of years, or lifetime of the power plant, and $R_t$ is the net cash flow of the year $t$:
\begin{equation} \label{eq:npv_eq}
NPV(t, N) = \sum_{t=0}^{N}\frac{R_t}{(1+t)^t}.
\end{equation}

$R_t$ takes into account the expected carbon price ten years into the future. Generator companies use a linear regression based on historical data to project the carbon price. With an increasing carbon price, the $R_t$ reduces for high carbon-emitting power plants, therefore making low carbon options more desirable. As demand in ElecSim is exogenous, demand does not change with respect to price and is assumed to be perfectly price inelastic. 

The yearly income for each power plant is estimated for each generation company by running a merit-order dispatch electricity market ten years into the future. However, the expected cost of electricity ten years into the future is uncertain. We, therefore, use the reference scenario projected by BEIS and use the predicted costs of electricity calibrated by Kell \textit{et al} \cite{DBEIS2019, Kell2020}. The agents predict the future carbon price by using a linear regression model.

\subsection{Optimization}

In this section, we detail the optimization approach taken. We modify the carbon tax each year, as we believe this is the most likely process taken by governments, giving generator companies and consumers the ability to understand market conditions during each year.

\label{ssec:optimization}
\subsubsection{Non-parametric carbon policy}
\label{sssec:non_parametric_strategy}
The optimization approach has two stages. First, we initialize the population of the NSGA-II algorithm $P_0$ with 18 attributes. These correspond to a separate carbon tax for each year, shown by Equation \ref{eq:eighteen_degrees_freedom}:
\begin{equation}
\label{eq:eighteen_degrees_freedom}
    P_0=\{a_1,a_2,\ldots,a_{18}\}, 0\leq a_y\leq 250,
\end{equation} 

\noindent where $P_0$ is the first population, $a_y$ is the attribute or carbon price in year $y$ and $a_1$ is the carbon price in year 1, $a_2$ the carbon price in year two and so forth. The constraints of the algorithm are that each of the carbon prices are bound between the values of \textsterling$0$ and \textsterling$250$. This provides the optimization algorithm with the highest degree of freedom. The value \textsterling$250$ was chosen due to the relative costs of electricity, where \textsterling$250$ would be the upper bound for the cost of electricity. This high degree of freedom enables a high number of strategies to be trialled due to its non-parametric nature. This, however, comes with a large search space requiring a large number of iterations.

 \subsubsection{Linear carbon policy}
 \label{sssec:linear_carbon_strategy}
 To reduce the search space for the carbon strategy, we also trial a linear carbon strategy, of the form:
 \begin{equation}
     p_c=a_1y_t+a_2, -14 \leq a_1\leq 14, 0 \leq a_2\leq 250,
 \end{equation}
 \noindent where $p_c$ is the carbon price, $y_t$ is the year, $a_1$ is the gradient or first attribute and $a_2$ is the intercept or second attribute. The constraints of the optimisation problem are that $a_1$ is bound by $-14$ and $14$, and $a_2$ by 0 and 250. These bounds are chosen to ensure that the carbon price does not exceed ${\sim}$\textsterling500 in the year 18 (2035) and is greater than about -\textsterling250, as well as ensuring that the carbon tax in the first year is greater than \textsterling0 but smaller than \textsterling250. The bounds for $a_1$ was chosen to make the mathematics simpler, whilst remaining in range.

\section{results}
\label{sec:results}

In this section, we explore the results of the optimizations, the optimum carbon strategies and the resultant electricity mixes.

\subsection{Non-parametric carbon policy}
\label{sssec:result_non_parametric_strategy}

\begin{figure}
\centering
\includegraphics[width=0.72\linewidth]{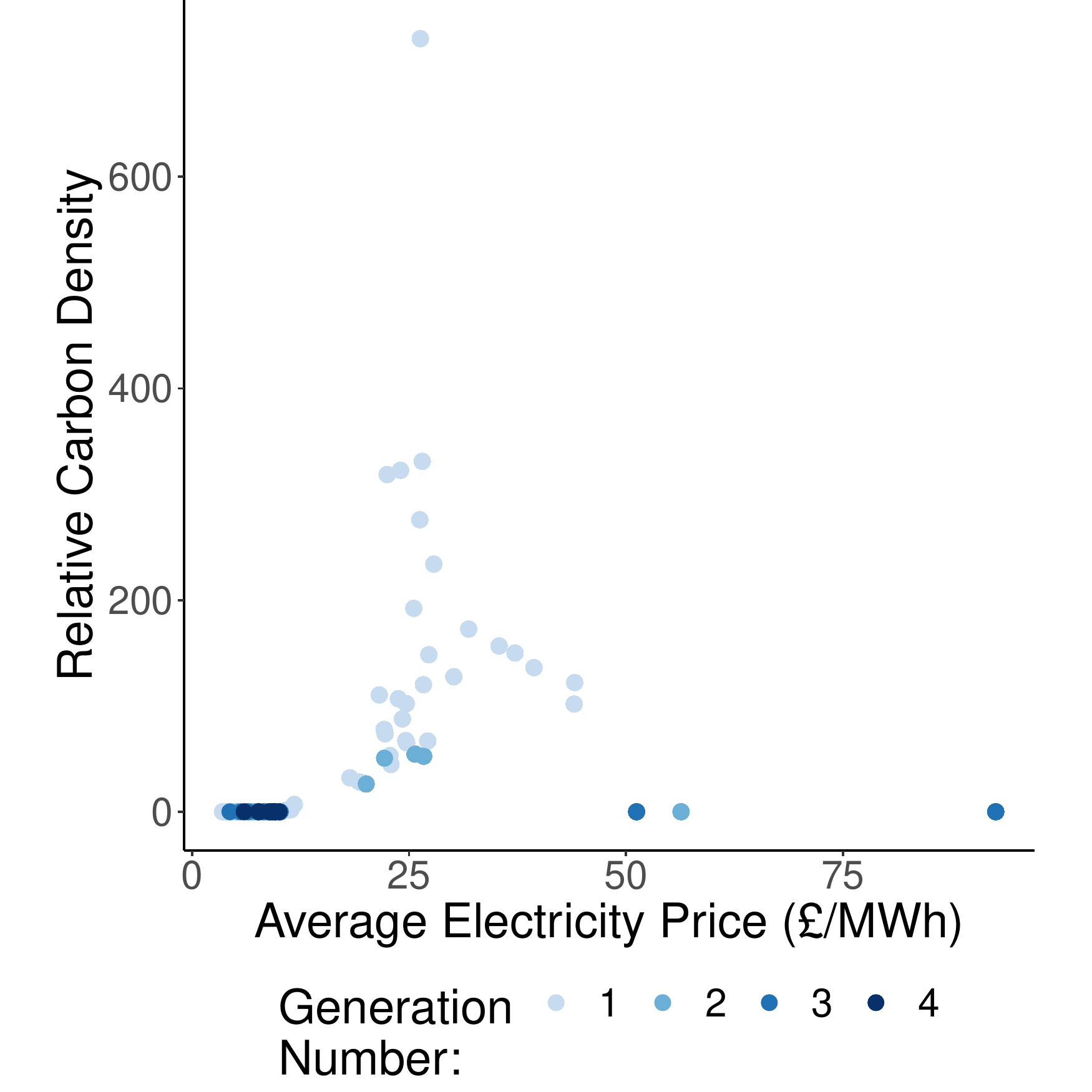}
\caption{Development of genetic algorithm rewards for non-parametric carbon tax policy results in 2035.}
\label{fig:free_points_ga_development}
%
\includegraphics[width=0.71\linewidth]{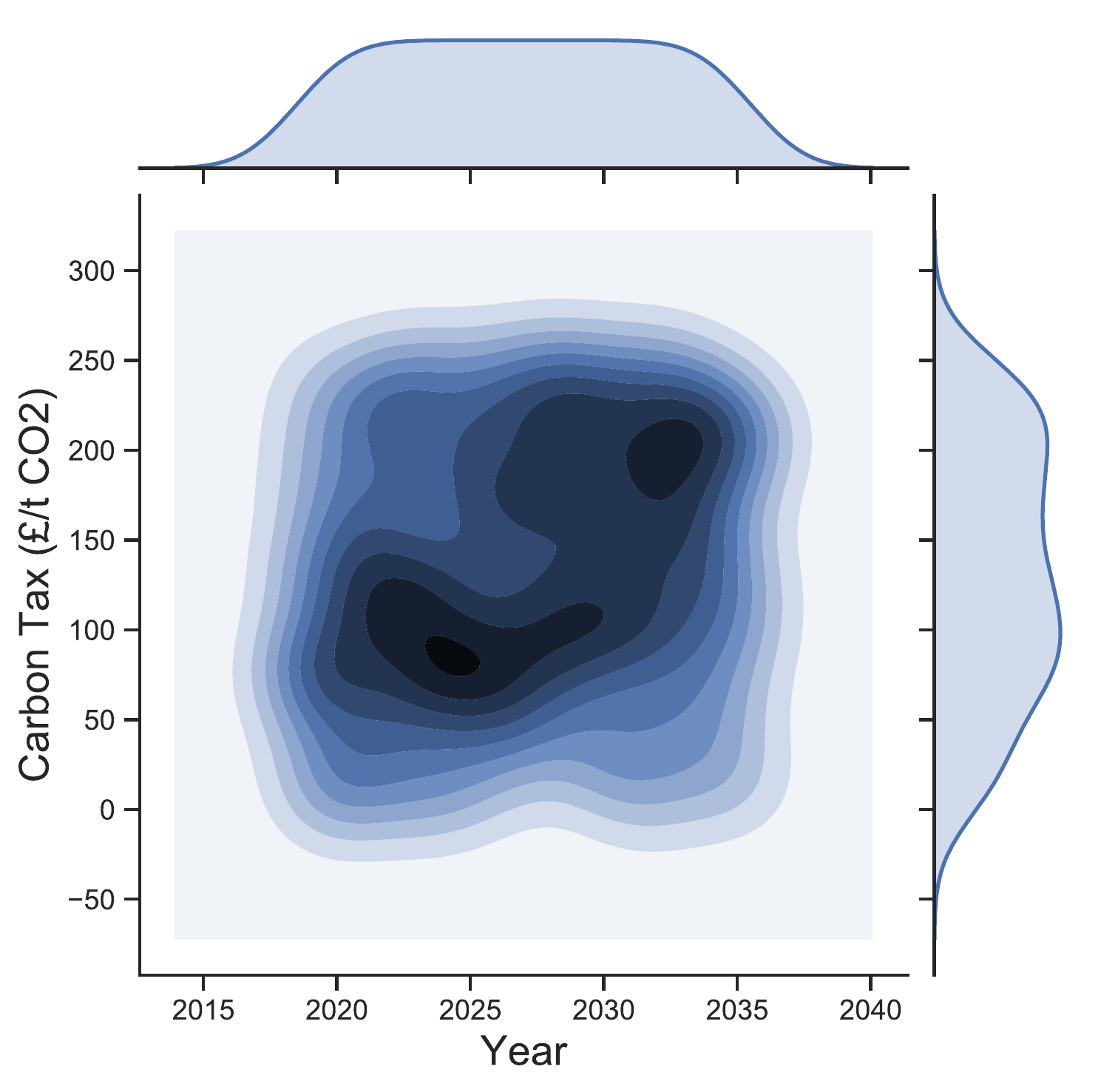}
\caption{Density plot of points with an average electricity price <\textsterling5/MWh for non-parametric carbon tax policy results in 2035.}
\label{fig:heatmap_of_free_points}
\end{figure}

Figure \ref{fig:free_points_ga_development} displays the development of the genetic algorithm against the rewards, relative carbon density and average electricity price. Darker colours display higher generation numbers. The first generation shows a widespread in relative carbon density and average electricity price. However, over successive generations, the solutions converge to a relative carbon density of 0 and an average electricity price under \textsterling10MW/h. 

Strikingly, the rewards of relative carbon density and average electricity price are not mutually destructive. This could be due to the low short-run marginal cost of renewable energy which reduces both electricity prices and carbon emissions~\cite{OMahoney2011}.

To understand the optimum carbon strategies, we visualized the parameters that produced the lowest average electricity prices in Figure \ref{fig:heatmap_of_free_points}. Specifically, we filtered for electricity prices under \textsterling5/MWh and displayed the results using a heat map. The darker colours represent a higher density of points. 

Figure \ref{fig:heatmap_of_free_points} displays a general trend, where carbon tax starts at ${\sim}$\textsterling100 until the year 2030, where it increases to ${\sim}$\textsterling200 by 2035. This may be due to the fact that a lower initial carbon tax of ${\sim}$\textsterling100 encourages investment in low-carbon technologies before the higher rate of ${\sim}$\textsterling200 comes into force. This higher rate of carbon tax would allow GenCos to outcompete higher carbon-emitting generators over the lifetime of the plants.


\subsection{Linear carbon policy}
\label{sssec:result_linear_carbon_strategy}

\begin{figure}
\centering
\includegraphics[width=0.70\linewidth]{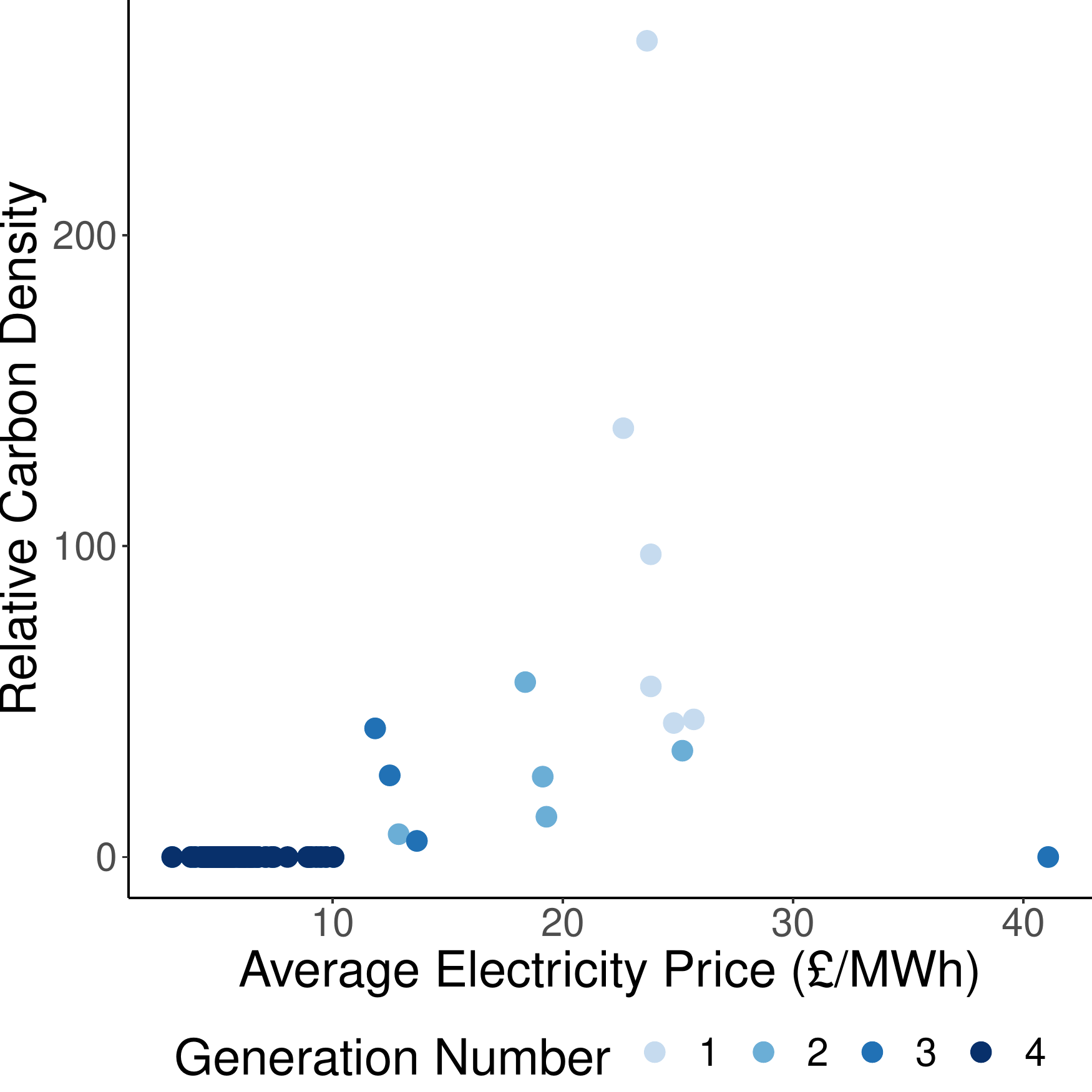}
\caption{Development of genetic algorithm rewards in 2035 for linear carbon strategy.}
\label{fig:linear_ga_development}
\end{figure}

Figure \ref{fig:linear_ga_development} displays the development of the genetic algorithm against the rewards: relative carbon density and average electricity price. Similarly to the non-parametric carbon policy shown in Figure \ref{fig:free_points_ga_development}, the first generation shows a wide spread of results. However, the spread is smaller than that of the linear carbon policy. This may be due to the fact that it is easier for the GenCos to predict the carbon policy, which increases confidence in the NPV calculations. The linear carbon policy also converges to a relative carbon density of 0, and an average electricity price smaller than \textsterling10MW/h.

\begin{figure}
\centering
\includegraphics[width=0.49\textwidth,]{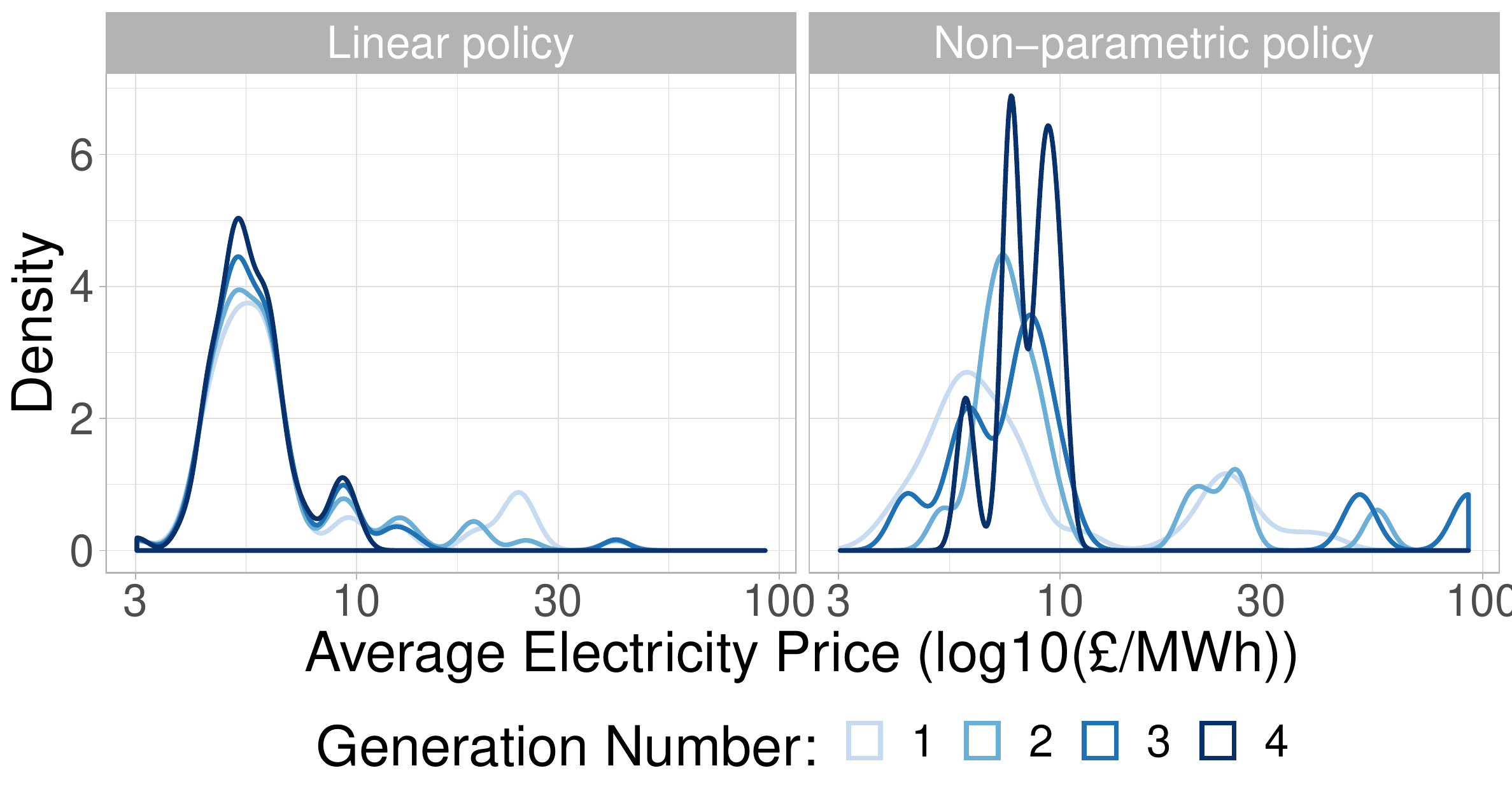}
\caption{Density plot of average electricity price in 2035 over generation number of genetic algorithm for both linear and non-parametric policy.}
\label{fig:comparison_of_distributions}
\end{figure}

Figure \ref{fig:comparison_of_distributions} compares the distributions of average electricity price for both techniques. Both methods show improvements as the number of generations of the genetic algorithm increase.  The linear policy, however, is able to more quickly converge to a low average electricity price, with a mode of ${\sim}$\textsterling5.4MW/h. The non-parametric policy has a number of poorer performing parameters, and Generation Number 4 has a bimodal distribution, with a mode of ${\sim}$\textsterling6.3MW/h.

\begin{figure}
\begin{subfigure}[h]{0.6\linewidth}
\includegraphics[width=\linewidth]{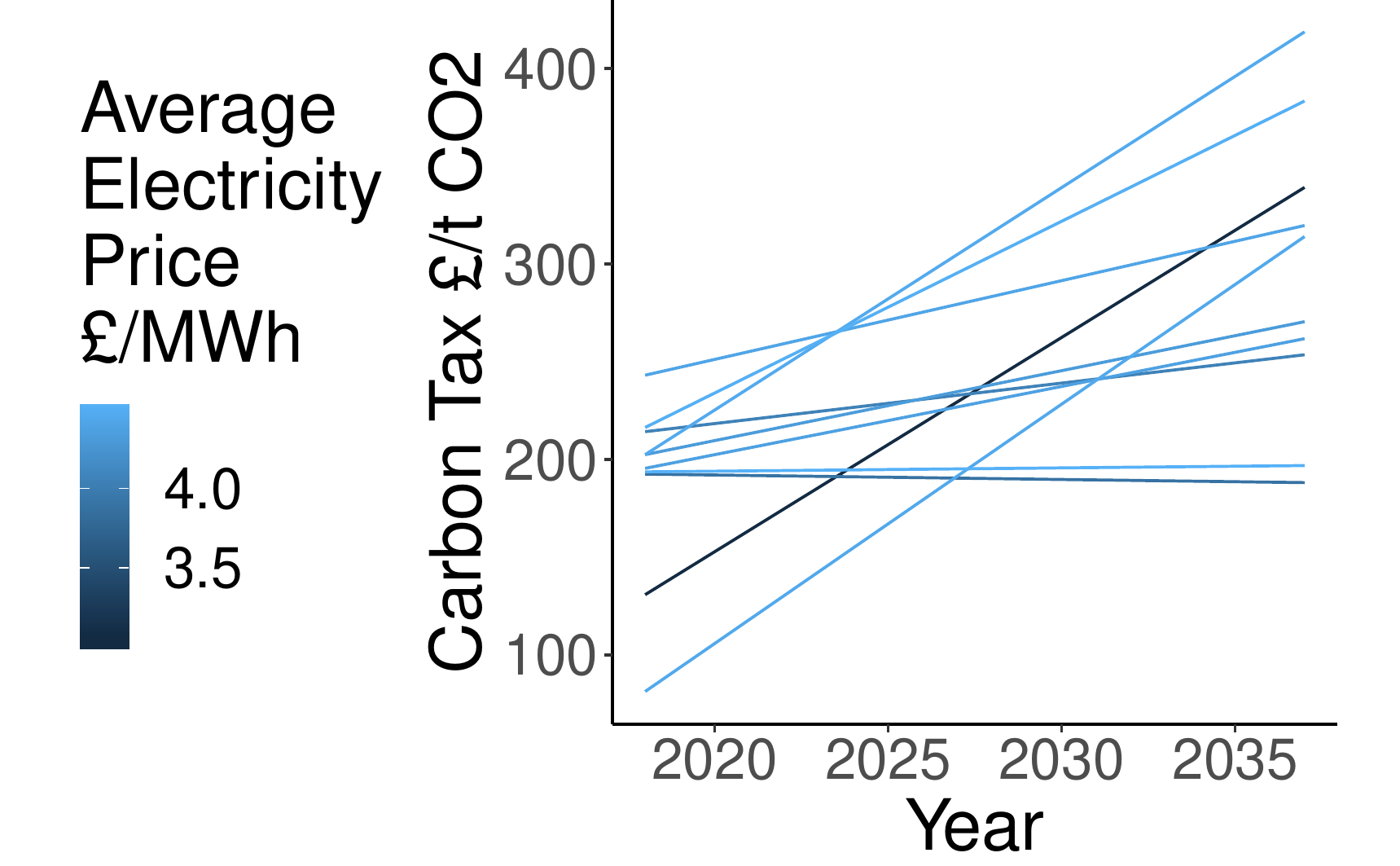}
\caption{All carbon policies.}
\label{fig:linear_actual_pdcs}
\end{subfigure}
\hfill
\begin{subfigure}[h]{0.39\linewidth}
\includegraphics[width=\linewidth]{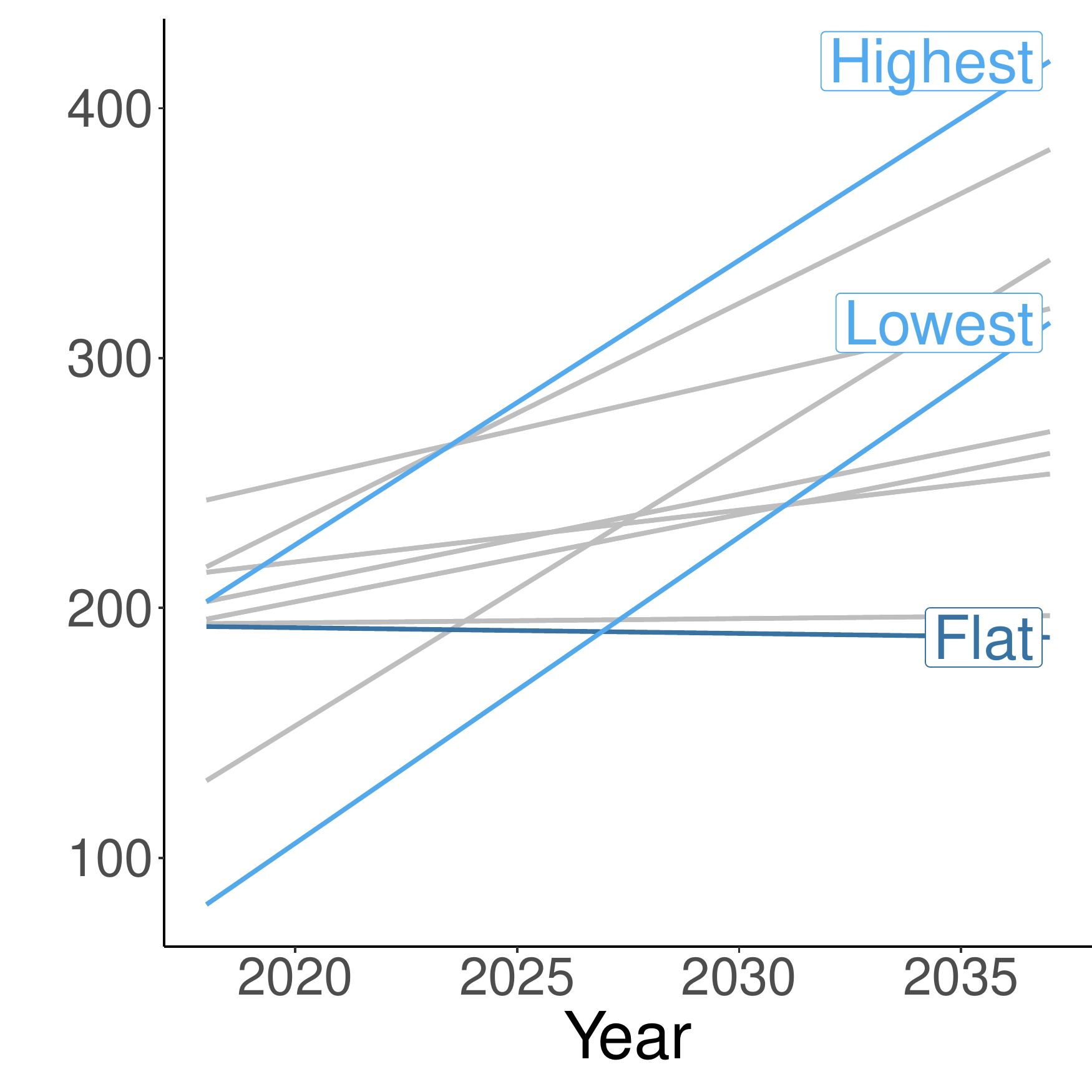}
\caption{Highlighted policies.}
\label{fig:highlighted_linear_actual_strategies}
\end{subfigure}%
\caption{Linear carbon policies under \textsterling4.5MW/h visualised.}
\end{figure}

Figure \ref{fig:linear_actual_pdcs} displays the linear carbon policies which had an average electricity price under \textsterling4.5MW/h. There is no single `optimum' carbon policy; a range of policies are able to achieve low carbon and a low average electricity price.

We explore the electricity mix generated of three different strategies shown in Figure \ref{fig:highlighted_linear_actual_strategies}. We selected the highest, lowest, and the lowest flat carbon strategy to show a range of possible strategies.

%

\begin{figure}
\centering
\includegraphics[width=0.50\textwidth]{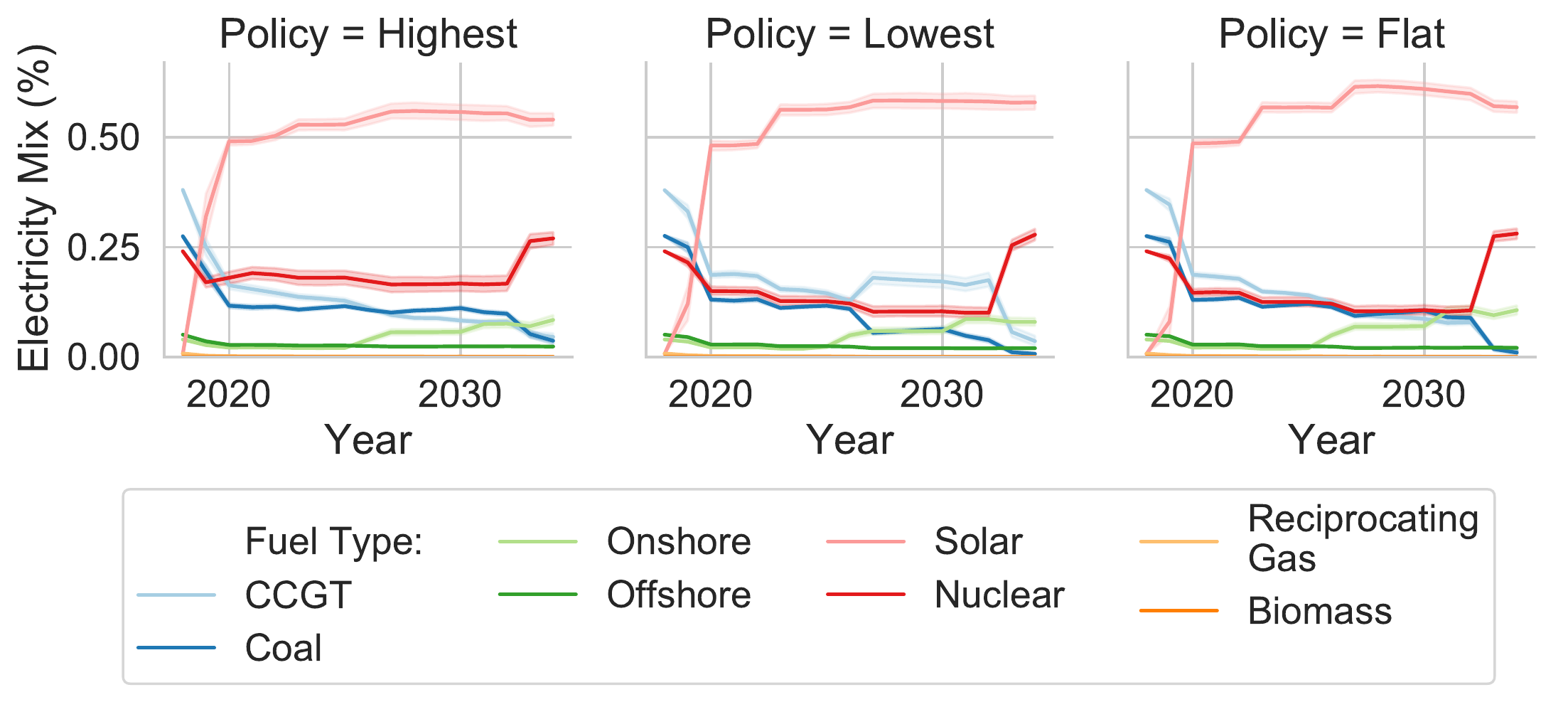}
\caption{Electricity mixes under selected linear carbon policies.}
\label{fig:best_electricity_mixes_facet}
\end{figure}

Figure \ref{fig:best_electricity_mixes_facet} displays the generated electricity mixes for each of the selected strategies. To generate these images, we ran 80 scenarios to capture the variability between scenarios. 

Whilst there does not seem to be a significant difference between scenarios, with solar providing ${\sim}60\%$ of the electricity mix by 2035, there is an observable difference with the other generator types.

The `highest' carbon strategy exhibits a higher uptake in nuclear, possibly due to the fact that nuclear becomes more competitive when compared to coal or gas. The `lowest' carbon strategy shows a higher uptake in Combined Cycle Gas Turbines (CCGT) during the years of 2026 to 2031 as it outcompetes nuclear. The `flat' carbon policy shows a higher percentage of solar energy than any of the other scenarios, albeit with a lower percentage of nuclear. Onshore wind is shown to be consistent for these scenarios.

\section{Conclusion}
\label{sec:conclusion}

In this paper, we have demonstrated that it is possible to use the genetic algorithm technique NSGA-II to optimize carbon tax policy using an electricity market agent-based model. 

We trialled a non-parametric carbon policy by allowing the genetic algorithm to optimize a carbon price for each year. These results showed us that a linear carbon tax might be appropriate. We then used a linear model as a carbon tax policy to reduce the total number of parameters for the genetic algorithm to optimize.  

We were able to show that a range of linear carbon taxes were able to achieve both low average electricity price and a relative carbon intensity of zero in 2035. By exploring three different carbon tax policies, we saw that ${\sim}$60\% of electricity consumption in the UK would be provided by solar. The difference between these `optimal' carbon tax policies was largely shown by competition between CCGT, coal and nuclear.

This was largely due to the low short-run marginal cost of solar and nuclear energy, which means that they are often dispatched ahead of the fossil-fuel based generators. CCGT and coal, however, are useful for filling demand when there is low solar irradiance.

In future work, we would like to try additional scenarios with varying future generation costs and calculate a sensitivity analysis to carbon taxes. In addition to this, we would like to model the uncertain reactions by consumers and generation companies with regards to carbon taxes. The linear carbon tax approach is an introductory approach which can be expanded upon.

\begin{acks}
This work was supported by the Engineering and Physical Sciences Research Council, Centre for Doctoral Training in Cloud Computing for Big Data [grant number EP/L015358/1].
\end{acks}
\clearpage
\bibliographystyle{ACM-Reference-Format}
\bibliography{library,custombibtex}

\appendix

\end{document}